\documentclass[prb,superscriptaddress,floatfix,twocolumn,showpacs,amsfonts,amsmath,amssymb]{revtex4}
\usepackage{graphicx} 
\usepackage{dcolumn} 
\usepackage{bm} 
\usepackage{color}

\begin{document}
\title{Nature of the spin liquid in underdoped cuprate superconductors}

\author{Y. A. Kharkov}
\affiliation{School of Physics, University of New South Wales, Sydney 2052, Australia}
\author{O. P. Sushkov}
\affiliation{School of Physics, University of New South Wales, Sydney 2052, Australia}
\date{\today}

\begin{abstract}

In the present work we address a long standing problem of the magnetic ground state and 
magnetic excitations in underdoped cuprates. 
Modelling cuprates by the extended $t-J$ model we show that there is a hidden dimensionless parameter $\lambda$ which drives magnetic
criticality at low doping $x$. Hence we derive the zero temperature 
$\lambda-x$ phase diagram of the model. It is argued that all  underdoped cuprates are 
close to the quantum tricritical point $x=0$, $\lambda=1$.
The three phases ``meet'' at the tricritical point: (i) N\'eel antiferromagnet, (ii) spin spiral
with antinodal direction of the spiral wave vector, (iii) algebraic spin liquid. 
We argue that underdoped cuprates belong either to the spin 
liquid phase or they are on the borderline between the spin liquid and the spin spiral.
We calculate the energy position $E_{cross}$ of the inelastic neutron scattering response
maximum at ${\bm q}=(\pi,\pi)$ and compare our results with experiments. We also explain softening of magnons
in the intermediate regime observed in inelastic neutron scattering.

\end{abstract}

\maketitle

\section{Introduction}
It is widely believed that an understanding of the nature of magnetic ground state and spin excitations in cuprates is crucial for 
resolving the problem of high Tc superconductivity.
The most striking physics arises in hole doped cuprates in the regime of low doping, where exotic phase transitions between distinct magnetic states take place.
Intricate details of doping driven transitions remain elusive and lack a unifying picture.
There are two major cuprate families, La$_{2-x}$Sr$_x$CuO$_4$ (LSCO) and YBa$_2$Cu$_3$O$\rm _{6+y}$ (YBCO) 
that are best experimentally studied in the low doping regime.
For a review of experimental data on magnetic excitations in these compounds see Ref.\cite{Fujita12} and also Refs.\cite{Stock08,Hinkov08,Haug10}.
While there are numerous material specific details (dependent on the degree of disorder, number of 
CuO$_2$ planes, oxygen chains, etc.), 
the most prominent and generic phenomenological observations can be summarized as follows.
(i) Commensurate antiferromagnetic (AFM) phase persist at very low doping, 
(ii) An intermediate state historically called the ``spin glass'' state arises in the doping window from a few per cent to about 10\%. The spin glass phase is characterized by very small static or quasi-static magnetic moments. 
(iii) At higher values of doping the static magnetic moment vanishes.
(iv) Magnetic response in the magnetically disordered phase is always incommensurate and manifests the famous ``hourglass'' dispersion.
(v) The onset of superconductivity upon increasing of doping always occurs in the ``spin glass'' phase.

\begin{figure}[h!]
\includegraphics[width=0.235\textwidth]{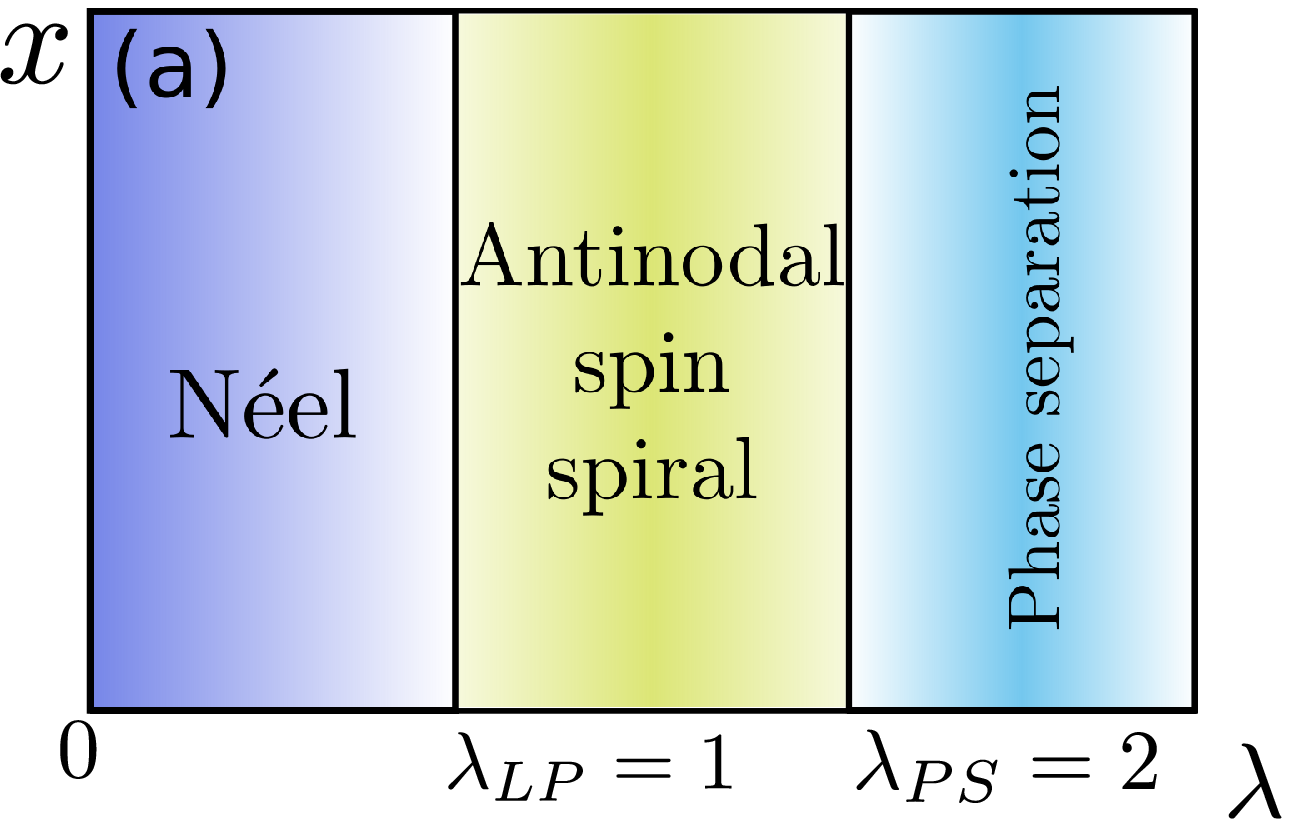}
\includegraphics[width=0.235\textwidth]{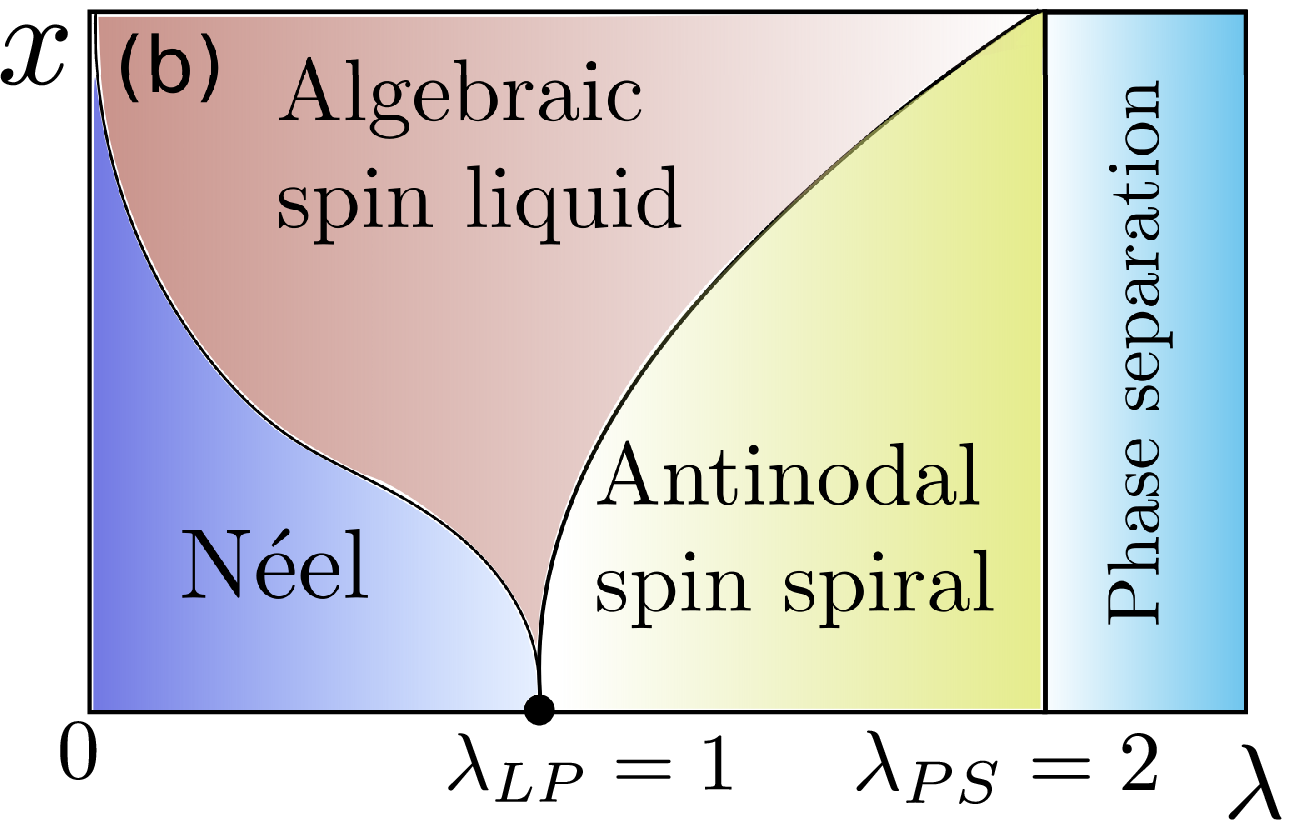}
\caption{Zero temperature phase diagram of a lightly doped extended $t-J$ model. 
(a) Classical phase diagram.\cite{Chubukov95}
(b) Quantum phase diagram. Strong quantum fluctuations in the vicinity of the Lifshitz point ($\lambda_{LP}$) result in a new algebraic spin liquid phase.
}
\label{F1}
\end{figure}

On the theoretical side it is widely accepted that the most important low energy physics of cuprates is described by the
extended $t-J$ model \cite{Anderson87, Emery87, Zhang88}.
Magnetic phase diagram of the $t-J$ model at the classical mean-field level, i.e. disregarding quantum fluctuations of spins,  
is well understood \cite{Shraiman90,Chubukov95}. Besides doping $x$, another important parameter is  $\lambda\propto g^2m^*$, where $g$ is the hole-magnon interaction constant and $m^*$ is the hole's effective mass. In a lightly doped $t-J$ model holes always form
small pockets near four nodal points in the Brillouin zone ${\bm k_0}=(\pm\pi/2,\pm\pi/2)$ and ${\bm k_0}=(\pm\pi/2,\mp\pi/2)$,  and $m^*$ describes curvature of the holon dispersion near the minima points.
The explicit relation of $\lambda$ to parameters of the extended $t-J$ model was derived in Ref.\cite{Sushkov04} and
will be specified later.
The zero temperature $\lambda-x$ mean-field phase diagram of the model is shown in Fig.\ref{F1}a.
The Lifshitz point at $\lambda_{LP}=1$ (Lifshitz  line) separates two phases: (i) the N\'eel phase at $\lambda<1$ and (ii) the static spin spiral phase at $1<\lambda<2$.
The direction of the spin spiral is always antinodal, i.e. ${\bf Q} = (Q,0)$ or 
${\bf Q} = (0,Q)$ and the wave vector of the spiral
scales linearly with doping, $Q \propto x$, Ref. \cite{Shraiman90}
When further increasing the coupling parameter $\lambda$ the system becomes unstable towards phase separation at 
$\lambda_{PS}=2$, Ref. \cite{Chubukov95}
A possibility of a noncoplanar state at $1 < \lambda < 2$ has been also considered \cite{Chubukov95}, however, the noncoplanar phase was ruled out in favour of the spin spiral state, see Ref.\cite{Milstein08}. 

In the mean field paradigm resulting in the phase diagram of the $t-J$ model in Fig. \ref{F1}a, 
quantum fluctuations of spins are completely ignored. 
On the other hand,  in the vicinity of the Lifshitz point quantum fluctuations are strongly enhanced and can lead to quantum phase transitions.
Some thirty years ago  Ioffe and Larkin considered a seemingly unrelated problem\cite{Ioffe88} of a Lifshitz transition in a two-dimensional (2D) frustrated antiferromagnet
(nonitinerant) between the collinear AFM phase and the spin spiral phases. Ioffe and Larkin showed that quantum 
fluctuations necessarily lead to a development of a gapped spin liquid phase in the vicinity of the 
Lifshitz point.
A frustration by itinerant fermions is very different from that in nonitinerant systems.
Nevertheless, in this work we show that quantum fluctuations in the $t-J$ model in the vicinity of the classical Lifshitz point also leads to the spin liquid phase
due to the mechanism similar to that by Ioffe and Larkin.
Hence, the classical Lifshitz line shown in Fig.\ref{F1}a expands to  a finite spin liquid region
shown in Fig.\ref{F1}b.
The endpoint of the classical Lifshitz line at $x=0$ becomes a quantum tricritical Lifshitz point.

In the present work we calculate the phase diagram, analyze properties of the spin liquid phase,
and compare our results with experimental observations for cuprates.
We argue  that cuprates belong to a relatively narrow vertical band near
$\lambda \approx 1$ in the phase diagram  Fig.\ref{F1}b.
In our analysis we consider the single layer model in the absence of disorder. Therefore our results are applicable to
cuprates at doping $x \gtrsim 5\%$. At doping lower than 5\%
the spin spiral physics in LSCO is driven by disorder
\cite{Sushkov2005,Luscher2007}, and in YBCO the physics is driven by the bilayer character of the compound \cite{sushkov2009}.
The spin liquid in the $t-J$ model, besides some similarities, has many differences from the Ioffe-Larkin
spin liquid in frustrated magnets. The most noticeable qualitative differences are (i) magnetic response in the spin liquid phase in the $t-J$ model has a finite spectral weight at low energies (magnetic pseudogap), in contrast to a fully gaped magnetic response in the Ioffe-Larkin case.
(ii)
The decay of spin-spin correlation with the distance is different in the two cases. In the Ioffe-Larkin spin liquid the correlator
decays exponentially with distance \cite{Kharkov18}. On the other hand, in the $t-J$ model
spin liquid is algebraic and the correlator decays as $1/r^{3}$.

\begin{figure}[h!]
\includegraphics[width=0.34\textwidth]{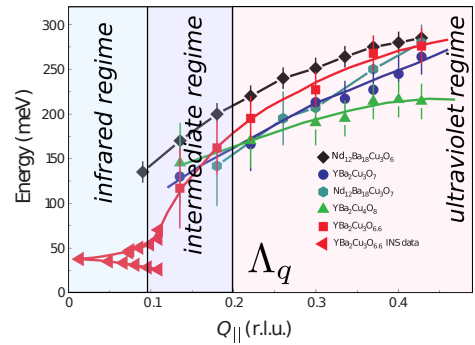}
\caption{Magnetic dispersion along the $(1,0)$ direction. Points show combined data on resonant 
inelastic X-ray scattering  and inelastic neutron scattering
in NdBCO and YBCO at $T=15$K, Ref. \cite{LeTacon11}. 
Vertical lines separate three different regimes that we call ``infrared regime'', ``intermediate regime'', and
``ultraviolet regime''.
}
\label{fig:RIXS_magnon}
\end{figure}

Following Refs.\cite{Shraiman90,Milstein08} we rely on 
quantum field theory formalism.
Interestingly, even experimental data indicates that the field theory is a very natural approach  to the problem.
In Fig.\ref{fig:RIXS_magnon} we present  magnetic dispersion along the $(1,0)$ crystal axis 
taken from Ref. \cite{LeTacon11}. The figure shows combined data on resonant 
inelastic X-ray scattering  and inelastic neutron scattering.
The data demonstrates three  distinct regimes separated in Fig.\ref{fig:RIXS_magnon} by vertical lines.
In the ``ultraviolet regime'' the dispersion only very weakly depends on doping,  practically doping 
independent. The independence is consistent with high temperature NMR data \cite{Slichter}.
In the ``intermediate regime'' there is a significant softening of the magnon dispersion with doping and the most 
dramatic doping dependence takes place in the ``infrared regime''.
We set the ultraviolet cutoff for the field theory $q\approx \Lambda_q$  that is the upper
edge of the ``intermediate regime'' as shown in Fig.\ref{fig:RIXS_magnon}.
The value of the cutoff indicated by the data is $\Lambda_q\sim 0.2(r.l.u) \sim 1.2/a$, where $a=3.81 \AA$ is the lattice spacing of the square CuO$_2$ plane.
In the main text we will determine the value of $\Lambda_q$ theoretically and show that  $\Lambda_q$ is
independent of doping.
The spin wave theory works well at $q > \Lambda_q$, moreover in this regime the field theory is not
valid and only the spin wave theory is applicable.
On the other hand, the magnon dispersion is linear in $q$  at $q \lesssim \Lambda_q$ and this justifies
applicability of the field theory.
The crossover energy scale between the ``intermediate regime'' and the ``infrared regime'' depends on doping
and the change of the regime is related to the energy $E_{cross}$ discussed in 
the experimental review in Ref. \cite{Fujita12}. We calculate values of $E_{cross}$ for different values of doping  and compare our results with data.
In the low doping limit, $x\to 0$, the size of the ``infrared'' domain shrinks to zero.
In our analysis, besides already mentioned publications, we use some ideas from 
Refs.\cite{Sachdev1994,sushkov1996,Onufrieva17,Chatterjee17}.

The paper is organized as follows. In Section \ref{sec:t-J} we review the  procedure for the reduction of the extended $t-J$ model to the quantum field theory.
The new point compared to already published results is the doping dependence of magnon speed.
In Section \ref{sec:HEP} 
we evaluate parameters of the field theory, calculate the  dependence (reduction) of magnon speed on doping using self consistent Born approximation and compare our predictions with 
inelastic neutron scattering data.
In Section \ref{sec:phase_diagr} we explain central ideas of the paper.
Here we discuss the theory of the  quantum Lifshitz transition driven by a coupling between spin excitations and low energy 
fermionic modes. In the same Section we present a magnetic phase diagram and derive properties of the new spin liquid phase. 
Here we separately consider a simple case of circular holon Fermi pockets and  more realistic case of elliptic pockets. 
In Section \ref{uv} we discuss the ultraviolet cutoff for the field theory, provide quantitative estimates for the ``Lindemann criterion'' of quantum melting. In the same section
we calculate $E_{cross}$ and compare it with the experimental data. Furthermore, we numerically evaluate the phase boundaries in zero temperature phase diagram.
In Section \ref{sec:correl} we consider the equal time  spin-spin correlator and demonstrate the
algebraic decay.
Finally, we summarize our results in Section \ref{sec:discuss}.

%
%
%

\section{Low energy limit of the extended $t-J$ model: quantum field theory}\label{sec:t-J}
\begin{figure}[h!]
\includegraphics[width=0.2\textwidth]{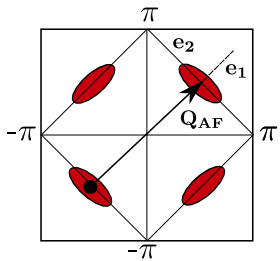}
\caption{Holon's Fermi pockets in underdoped cuprates. }
\label{FS1}
\end{figure}

The Hamiltonian of the extended $t-J$ model reads \cite{Anderson87, Emery87, Zhang88}
\begin{eqnarray}\label{eq:Ham_tJ}
H = -t\sum_{\langle ij \rangle} c^\dag_{i, \sigma} c_{j, \sigma} -t'\sum_{\langle \langle ij \rangle\rangle} c^\dag_{i, \sigma} c_{j, \sigma} - \nonumber \\
t''\sum_{\langle \langle \langle ij \rangle\rangle\rangle} c^\dag_{i, \sigma} c_{j, \sigma} + J \sum_{\langle i,j \rangle} \left[ \bm S_i\cdot \bm S_j - \frac{1}{4}N_i N_j \right],
\end{eqnarray}
where $c^\dag_{i\sigma}$ ($c_{i\sigma}$) is the creation (annihilation) operator for an electron with spin $\sigma=\uparrow,\downarrow$ at Cu site $i$; the operator of electron spin reads $\bm S_i = \frac{1}{2} c^\dag_{i\alpha} \bm\sigma_{\alpha\beta} c_{i\beta}$. The electron number density operator is $N_i = \sum_\sigma c^\dag_{i\sigma} c_{i\sigma}$, where $x$ is the hole doping, so that the sum rule $\langle N_i \rangle = 1-x$ is obeyed.
In addition to Hamiltonian (\ref{eq:Ham_tJ}) there is the no double occupancy constraint,
which accounts for a strong electron-electron on-site repulsion. 
Values of parameters slightly vary between different compounds. 
Typically $J \approx 125\, meV$ and the hopping integrals are 
$t \approx 390meV\approx 3J $, $t' \approx −90meV \approx -0.7J $, 
$t'' \approx 80meV \approx 0.6J $ , see e.g. Ref.\cite{andersen95}
The Fermi surface of a lightly doped extended $t-J$ model consists 
of Fermi pockets shown in Fig.\ref{FS1} and centered
at the nodal points $\mathbf{k}_{0}=(\pm \pi /2,\pm \pi /2)$, and $\mathbf{k}_{0}=(\pm \pi /2,\mp \pi /2)$.
The hole dispersion can be approximately calculated using a self-consistent Born approximation (SCBA), that is well known to be very reliable for the single hole problem in the $t-J$ model. The single hole dispersion
can be parametrized as \cite{Sushkov1997}
\begin{eqnarray}
\label{disp}
&&\epsilon_{\bm k}=\beta_1(\gamma_{\bm k}^+)^2+\beta_2(\gamma_{\bm k}^-)^2  , \ \ 
\gamma_{\bm k}^{\pm}=\frac{1}{2}(\cos k_x \pm \cos k_y),\nonumber\\
&&\epsilon_{\bm k}\approx \beta_1\frac{p_1^2}{2}+\beta_2\frac{p_2^2}{2}.
\end{eqnarray} 
Hereafter we set the lattice spacing equal to unity, $a= 3.81\thinspace \AA$ $\,\rightarrow $
\thinspace 1. The second line in Eq.(\ref{disp}) corresponds to the quadratic expansion of the fermion dispersion along the principle
axes of the Fermi surface ellipse, Fig.\ref{FS1},  ${\bf p}={\bf k}-\mathbf{k}_{0}$, and Fermi energy is related to doping as
\begin{eqnarray}
\label{eF}
&&\epsilon_F\approx \pi\beta x, \nonumber\\
&&\beta=\sqrt{\beta_1\beta_2}=\frac{1}{m^*}.
\end{eqnarray}

Inverse effective masses $\beta_1$, $\beta_2$ can be calculated within the extended $t-J$
model in SCBA approximation and they significantly depend on $t'$ and $t''$, see Ref.\cite{Sushkov04}
At values of $t'$ and $t''$, corresponding to cuprates, the inverse effective mass is
$2J < \sqrt{\beta_1\beta_2} <2.5J$.
Hence, the effective mass of a hole is approximately 
twice the  electron mass, $m^* \approx 2m_e$.

While the $t-J$ model is the low energy reduction of the three band Hubbard model, the total energy
range in the $t-J$ model, $\Delta\epsilon \sim 8t \sim 24J\approx 3$eV, is still very large. 
On the other hand we are interested in the energy interval bounded by the top edge of
the intermediate regime in Fig.\ref{fig:RIXS_magnon}, $E \lesssim 150-200$meV.
Therefore, for our purposes it is quite natural to consider the low energy sector of $t-J$ model. 
The effective low energy Lagrangian was first 
derived in Ref.\cite{Shraiman90} with some important terms responsible for stabilty of the spin
spiral ground state missing. The full effective Lagrangian was derived in Ref.\cite{Milstein08}
This approach necessarily requires an introduction of two checkerboard sublattices, independent of  
whether there is a long range AFM order or the order does not exist.
The two checkerboard sublattices allow us to avoid a double counting of quantum states in the case
when spin and charge are separated.
A hole, which hereafter we call a holon, does not carry a spin, but it can be located at one of the
sublattices and this is described by the pseudospin $1/2$.
Due to the checkerboard sublattices the Brillouin zone coinsides with magnetic   
Brillouin zone (MBZ) even in the absence of a long range AFM order.
Therefore, there are four half-pockets in Fig.\ref{FS1} or two full pockets within MBZ.
Finally, the Lagrangian reads\cite{Milstein08}
\begin{eqnarray} 
\label{eq:LL}
{\cal L}&=&\frac{\chi_{\perp}}{2}{\dot{\vec n}}^2- 
\frac{\rho_s}{2}\left({\bm \nabla}{\vec n}\right)^2\\
&+&\sum_{\alpha}\left\{ \frac{i}{2}
\left[\psi^{\dag}_{\alpha}{{\cal D}_t \psi}_{\alpha}-
{({\cal D}_t \psi_{\alpha})}^{\dag}\psi_{\alpha}\right] - \psi^{\dag}_{\alpha}\epsilon_{\alpha}({\bf \cal P})\psi_{\alpha} \right.\nonumber\\
&+&\left. \sqrt{2}g (\psi^{\dag}_{\alpha}{\vec \sigma}\psi_{\alpha})
\cdot\left[{\vec n} \times ({\bm e}_{\alpha}\cdot{\bm \nabla}){\vec n}\right]\right\} \  .
\nonumber
\end{eqnarray}
Fermions (holons) are described by a spinor $\psi_\alpha$ with the pseudospin $1/2$, and the vector of staggered magnetization ${\bm n}$ normalized as $\bm n^2=1$ corresponds to localized spins at Cu sites.
The first line in (\ref{eq:LL}) is $O(3)$ nonlinear sigma model that describes spin dynamics, the second line is the Lagrangian for non-interacting holons. 
The long covariant derivatives in Eq. (\ref{eq:LL}) are defined as
\begin{eqnarray}
&\mathcal{ \bm P} = -i\bm \nabla + \frac{1}{2}\vec\sigma\cdot [\vec n \times \bm \nabla \vec n],\\
&\mathcal{D}_t = \partial_t + \frac{1}{2} \vec\sigma \cdot [\vec n \times  \partial_t \vec {n}].
\end{eqnarray}
The index $\alpha=1,2$ enumerates two full holon pockets in Fig. \ref{FS1}. 
The term in the bottom line in Eq. (\ref{eq:LL}) describes a coupling between holons and
the staggered magnetization. 
Pauli matrices $\bm\sigma$ in Eq. (\ref{eq:LL}) act on the holon's pseudospin and $\bm e_\alpha = 1/\sqrt{2}(1,\pm1)$ 
denotes a unit vector orthogonal to the face of the MBZ where the holon is located.

Lagrangian (\ref{eq:LL}) contains five parameters, $\chi_\perp$,  $\rho_s$,
$\beta_1$, $\beta_2$, and $g$. 
Parameters of a quantum field theory always depend on the energy/momentum scale and hence the values of the parameters are fixed at a particular normalization point.
We use the ultraviolet limit $\Lambda_q$ discussed in
the Introduction as the normalization point.  In the limit $x\to 0$ the $\sigma$-model
parameters $\chi_\perp$ and   $\rho_s$ coincide with that of the 2D Heisenberg model on the square lattice, 
$\chi_\perp=1/8J$, $\rho_s=J/4$ and the magnon speed 
\begin{eqnarray}
\label{c0}
c_0=\sqrt{\rho_s/\chi_{\perp}}=\sqrt{2}J .
\end{eqnarray}
The coupling constant is $g=Zt$, where $Z$ 
is the holon quaziparticle residue
calculated within the $t-J$ model \cite{Sushkov1997}. For  $t'$ and $t''$ corresponding to cuprates
and even for $t'=t''=0$ the coupling constant  is always close to $g\approx J$.

The most important parameter that drives magnetic quantum criticality in the model is the effective fermion-magnon coupling strength\cite{Milstein08}
\begin{equation}
\label{ll}
\lambda = \frac{2 g^2}{\pi \rho_s \sqrt{\beta_1 \beta_2}}.
\end{equation}
Lagrangian (\ref{eq:LL}) has been analyzed previously in a classical mean-field approximation.
The phase diagram obtained in this approximation is shown in Fig.\ref{F1}a.
The collinear AFM state is stable at $\lambda< 1$. At $\lambda > 1$ the 
spin spiral is developing, the wave vector of the spiral
depends linearly on doping, $Q \propto x$, Ref. \cite{Shraiman90}
The direction of the spiral wave vector is antinodal, i.e. ${\bf Q}\propto (1,0)$ or 
${\bf Q} \propto (0,1)$ and
at further increasing of $\lambda$ a phase separation instability is developing at 
$\lambda_{PS}=2$, Ref. \cite{Chubukov95,Milstein08}.
Taking the values of the field theory parameters corresponding to cuprates, as described in the previous
paragraph, the value of $\lambda$ is $1<\lambda<1.3$. \cite{Sushkov11}
In theory one can vary $\lambda$ arbitrarily. For example in the pure $t-J$ model,
$t'=t''=0$, the value of $\beta_2$ is very small and hence  $\lambda > 2$, the model
is unstabe with respect to the phase separation\cite{Sushkov04a}.
Within the extended $t-J$ model it is rather hard to make $\lambda$ significantly smaller than 1. 
For instance, using the set of the $t-J$ model parameters with an unreasonably high value of $t''$,
$t= 3J $, $t'=0 $, $t''=3J $ the SCBA approximation gives  
$\lambda \approx 0.7$. This set of parameters is unphysical.
For realistic parameters of cuprates $\lambda$ 
is close to unity and  probably  slightly higher than unity. We estimate the interval for the parameters of cuprates as
\begin{eqnarray}
\label{lll}
0.9 < \lambda  < 1.3.
\end{eqnarray}
While there is no experimentally available handle that would allow to directly tune parameter $\lambda$ in a given cuprate compound, parameter $\lambda$ is vital for the description of phase transitions between different magnetic states in cuprates.


\section{Softening of magnons in the ``intermediate'' regime}\label{sec:HEP}

Softening of magnons with doping in cuprates was observed in inelastic neutron scattering  long 
time ago, see Refs. \cite{Bourges1997,Lipscombe09,Vignolle07}, see also an experimental review in Ref.\cite{Fujita12}.
This phenomenon still lacks a theoretical explanation.
In this section we calculate the dependence of the field theory parameters on doping and as a
byproduct of this analysis we explain the softening.
The physics discussed in the present section concerns relatively high energies and it is
independent of the Lifshitz magnetic criticality that is driven by $\lambda$ and  is
discussed in subsequent Sections.  

\begin{figure}[h!]
\includegraphics[width=0.35\textwidth]{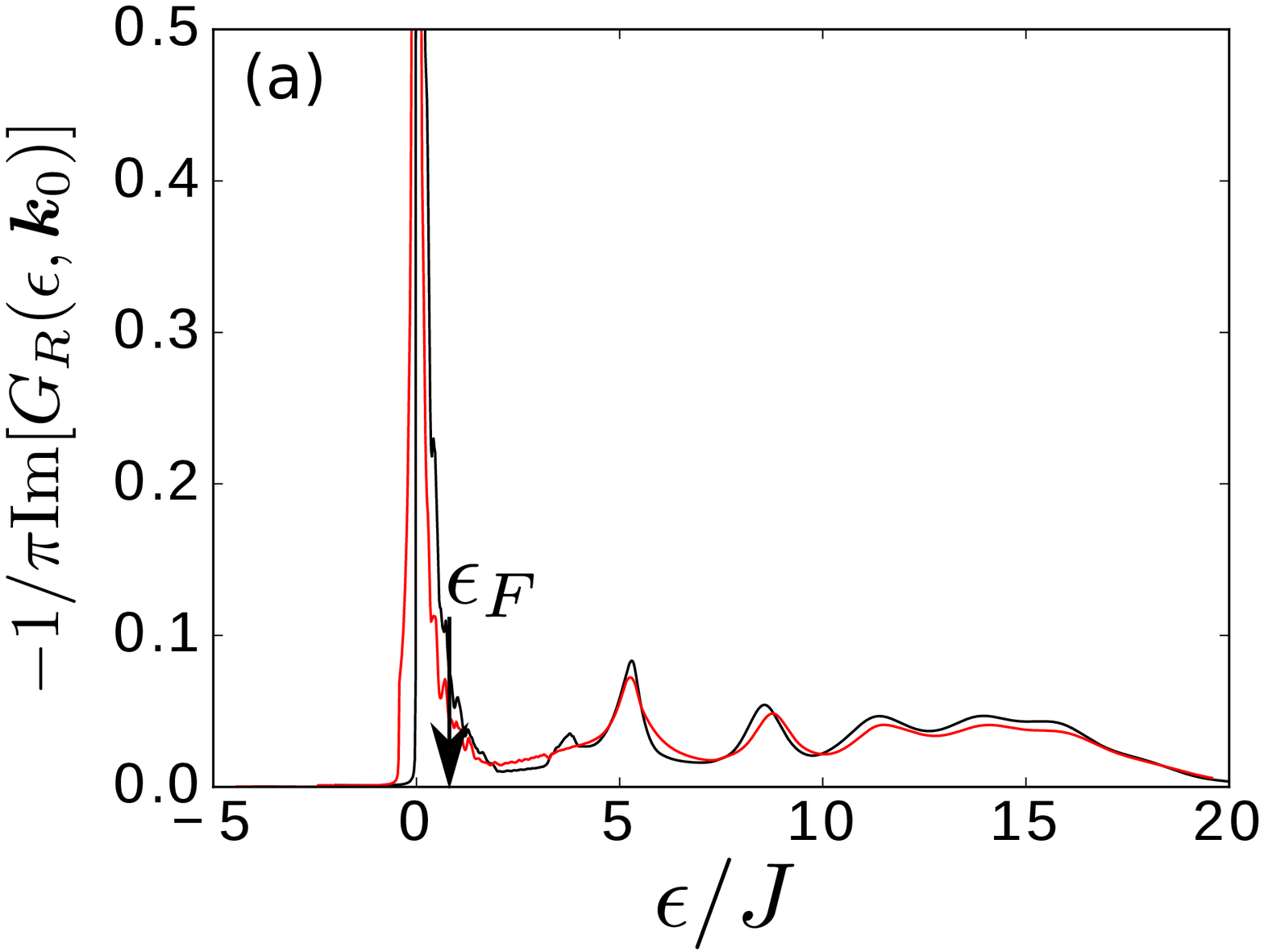}
\includegraphics[width=0.3\textwidth]{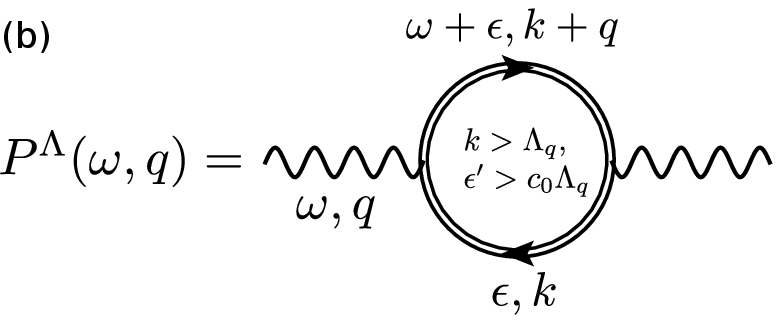}
\caption{
(a) Spectral density of holon's retarded Green's function at ${\bf k_0}=(\pi/2,\pi/2)$
calculated\cite{Sushkov1997} numerically in SCBA.
The black line corresponds to the set of parameters $t=3.1J$, $t'=-0.5J$, $t''=0.4J$, 
and the red line corresponds to $t=3.1J$, $t'=-0.8J$, $t''=0.7J$. The arrow shows the value of the Fermi energy for the doping interval $5\%\leq x\leq 15\%$.
(b) Magnon polarization operator $P^\Lambda(\omega, \bm q)$, the double line represents dressed  holon Feynman Green's function.
}
\label{hGF1}
\end{figure}

The single hole problem in the $t-J$ model was solved decades ago using SCBA, and we will skip all technical details of such calculations.
The spectral density of a single holon retarded Green's function 
\begin{equation}
G_R(\epsilon,{\bf k}) = -i \int dt d\bm r \, e^{-i\epsilon t + i\bm k \bm r} \langle T\{c_\uparrow^\dag(\bm r, t) c_\uparrow(0,0)\}\rangle
\end{equation}
 is plotted in Fig.\ref{hGF1}a.
The spectral density can be represented as
\begin{eqnarray}
\label{sfR}
-\frac{1}{\pi}Im [G_R(\epsilon,{\bf k})]= Z_{\bf k}\delta(\epsilon-\epsilon_{\bf k}) 
+ \rho_{\bf k}(\epsilon).
\end{eqnarray}
Here $Z_{\bf k}$ is the holon's quasiparticle residue, $\epsilon_{\bf k}$ is the holon's dispersion (\ref{disp}),
and $\rho_{\bf k}$ is the the incoherent ``tail''.
The incoherent tail stretches up to very high energies and is equal to the energy span of the $t-J$ model $\Delta\epsilon\sim 8t \sim 24J\approx 3$eV.
At small doping the spectral density of Feynman Green's function can be expresses in terms of
(\ref{sfR}), Ref.\cite{Chen13}
\begin{eqnarray}
\label{sfF}
&&-\frac{1}{\pi}Im [G_F(\epsilon,{\bf k})]= sgn(\epsilon-\epsilon_F)Z_{\bf k}\delta(\epsilon-\epsilon_{\bf k})
+ \rho_{\bf k}(\epsilon),\nonumber\\
&&sgn(z)=z/|z|,
\end{eqnarray}
where $\epsilon_F$ is Fermi energy (\ref{eF}).
As one can see from Fig.\ref{hGF1}a the incoherent part is negligible at energies below the Fermi energy, $\epsilon<\epsilon_F$.
The magnon polarization operator $P^{\Lambda}(\omega,{\bf q})$ 
is given by the fermionic loop shown in 
Fig.\ref{hGF1}b. The magnon Green's function defined on the local antiferromagnetic background 
directed along the $z$-axis reads
\begin{eqnarray}
\label{gfrl}
&&{\vec n}_{\perp}=(n_x,n_y,0)\nonumber\\
&&{\vec n}=(n_x,n_y,\sqrt{1-n_{\perp}^2})\nonumber\\
&&D(t,{\bf r})\delta_{\alpha\beta}=-i\langle T\left\{{ n}_{\perp, \alpha}(t,{\bf r}){ n}_{\perp, \beta}(0,0)\right\}\rangle.
\end{eqnarray}
The standard expression for magnon Green's function in a single loop approximation reads
$D(\omega, \bm q)=\chi_{\perp}^{-1}[\omega^2-c^2 \bm q^2-P(\omega,{\bf q})+i0]^{-1}$.
In what follows we will separately consider the magnon's Green's function in the two regimes corresponding to the two energy/momentum scales. (i) ``Ultraviolet'' scale,  $q\sim \Lambda_q\approx 1$,
$\omega \sim c_0\Lambda_q \sim J\sim 150$ meV, where all slow fluctuations related to magnetic criticality are irrelevant.
In this regime the magnon's propagator is
\begin{equation}
\label{gfrl}
D^\Lambda(\omega, \bm q)=\frac{\chi_{\perp}^{-1}}{\omega^2-c_0^2 \bm q^2-P^\Lambda(\omega,{\bf q})+i0},
\end{equation}
where we use the ``bare'' magnon speed $c_0=\sqrt{2}J$ and the polarization operator $P^\Lambda$ is shown in Fig. \ref{hGF1}b. In this regime in the polarization bubble $P^\Lambda(\omega, \bm q)$  only high energy particle-hole excitations with energies $|\epsilon_k-\epsilon_F| \gg \Lambda$  are accounted, that is emphasized by the superscript $\Lambda$. 
As we demonstrate in the present Section, the polarization operator $P^\Lambda$ is responsible for the reduction of the magnon speed with doping.
The second energy scale corresponds to (ii) ``intermediate''+``infared'' regimes which are presented in Fig. \ref{fig:RIXS_magnon}. 
The physics in the case (ii) is related to magnetic criticality, and  will be addressed in following sections.
The information about the ``ultraviolet'' physics is incorporated in the low energy physics (ii) via renormalized parameters of the Lagrangian (e.g. renormalized magnon speed $c$).

%
The fermion loop diagram $P^\Lambda$, shown in Fig.\ref{hGF1}b, contains a product
of the positive and negative frequency components of the fermion Feynman Green's function (\ref{sfF}).
So, there are two main contributions to the polarization operator: coherent-coherent $\propto Z_k Z_{k+q}$
($k<p_F$, $|{\bf k}+{\bf q}|>p_F$), and coherent-incoherent $\propto Z_k\rho_{k+q}$ ($k<p_F$, 
$|{\bf k}+{\bf q}|$ is arbitrary).
The first contribution is the most important one for the quantum critical physics in the ``infared regime''
and actually it cannot be calculated within the ``simplistic'' logic of this section.
The physics in the ``infared regime'' will be considered in the next section.
Luckily, the coherent-coherent contribution declines
 with energy and at the top edge
of the ``intermediate regime'', $\omega \sim c_0\Lambda_q$, this contribution is negligible.
On the other hand, the coherent-incoherent contribution is important everywhere including the top edge
of the ``intermediate regime''.
Finally, the ``incoherent-incoherent'' contribution $\propto \rho_k \rho_{k+q}$ is strongly suppressed, since the incoherent part $\rho_k$ is negligible below the Fermi energy, $\epsilon<\epsilon_F$.
 Hence, the polarization operator reads
\begin{eqnarray}
\label{poL1}
&&P^{\Lambda}(\omega, {\bf q})\approx 4 \sum_{\alpha=1,2}\int_{k < p_F}\frac{d^2 k}{(2\pi)^2}
({\tilde g}_{{\bf k},{\bf q}})^2\nonumber\\
&&\times
\int_{c_0\Lambda_q}^{\infty}dy 
\frac{Z_{\bf k}\rho_{{\bf k}+{\bf q}}(y)}{\omega-\epsilon^\alpha_{\bf k}-y} 
 \ .
\end{eqnarray}
Here $\alpha$ enumerates holon pockets, and the the holon-magnon vertex ${\tilde g}_{{\bf k},{\bf q}}$ is
related to $g_{{\bf k},{\bf q}}$ from Ref.\cite{Sushkov1997} as
\begin{eqnarray}
\label{v1}
{\tilde g}_{{\bf k},{\bf q}}=\sqrt{2\omega_q}g_{{\bf k},{\bf q}}=4\sqrt{2}t\sqrt{2\omega_q}
(\gamma_{\bf k} u_{\bf q}+\gamma_{\bf k+q} v_{\bf q})
\end{eqnarray}
The factor $\sqrt{2\omega_q}$ in the vertex ${\tilde g}_{{\bf k},{\bf q}}$ is due to a normalization. Here we use the standard quantum field theory normalization for the magnon field while
Ref.\cite{Sushkov1997} has used the Schr\"odinger equation normalizatioin.  
In the vicinity of a given Fermi pocket ${\bf k} \approx {\bf k}_0=(\pi/2,\pm\pi/2)$ and at $q< 1$
the vertex (\ref{v1}) reads
\begin{eqnarray}
\label{v2}
{\tilde g}_{{\bf k},{\bf q}}\approx  4t \sqrt{2J} q_{1,\alpha} \ ,
\end{eqnarray}
where $q_{1,\alpha}$ is the component of the momentum orthogonal to the face of the MBZ in this pocket.

\begin{figure}
\includegraphics[scale=0.27]{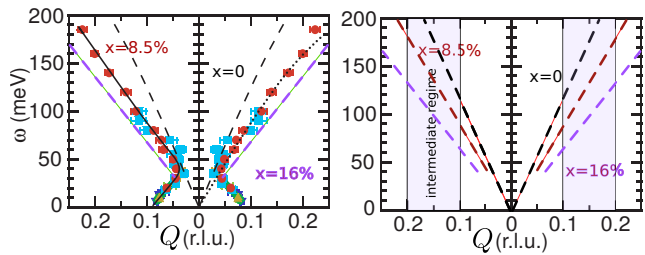}
\caption{Magnon dispersion in the ``intermediate regime'', $70\, meV < \omega < 200\, meV$,
Left panel presents experimental data  for $x=0$ (Ref.\cite{Coldea2001}),
$x=0.085$ (Ref.\cite{Lipscombe09}), and 
$x=0.16$ (Ref.\cite{Vignolle07})
(black, red and magenta dashed lines, respectively).
 Right panel: Theoretical magnon dispersion (\ref{swv}) for the same values of doping.
Shaded regions indicate the ``intermediate regime'' where the right and the left panels
should be compared.}
\label{soft}
\end{figure}

The incoherent component of the holon's Green's function remains approximately constant $\rho(y)\approx 
(1-Z_k)/8t$ in the energy interval $ 2J\lesssim\epsilon\lesssim 8t$, see Fig. \ref{hGF1}a.
The latter estimate for the incoherent part of the holon's spectral function
follows from the sum rule $Z_k+\int_{0}^{+\infty} d\omega \rho_k(\omega) = 1$. 
We also set $Z_k=J/t$.
Hence, using Eqs.(\ref{poL1}) and (\ref{v2}) we find
\begin{eqnarray}
\label{poL2}
P^{\Lambda}({\bf q})&\approx& -32t^2 q^2\left(J\int_{2J}^{\infty}
\frac{\rho(y)}{y}dy\right)\left[4\int_{k < p_F}\frac{d^2 k}{(2\pi)^2} Z_{\bf k}\right]
\nonumber\\
&\approx&-32t^2 x \ q^2 Z_{\bm k_0} \left(J\int_{2J}^{\infty}
\frac{\rho(y)}{y}dy\right) \approx \nonumber \\ 
&\approx& - 4 x J^2 q^2 \left( 1 -\frac{J}{t}\right) \ln \left({\frac{4t}{J}}\right).
\end{eqnarray}
For $t/J\approx 3$ this gives
\begin{equation}
\label{eq:P_Lambda}
P^{\Lambda}({\bf q}) \approx - 7 J^2 q^2 x.
\end{equation}
Direct numerical integration $\int dy \rho(y)/y $ in Eq. (\ref{poL2}) with the holon's Green's function 
plotted in Fig.\ref{hGF1}a 
results in $P^{\Lambda}({\bf q}) \approx - 8 J^2 q^2 x$, which is close to Eq. (\ref{eq:P_Lambda}).
Hence the magnon Green's function  (\ref{gfrl}) reads
\begin{eqnarray}
\label{gfrl1}
D^\Lambda(\omega, q)=\frac{\chi_{\perp}^{-1}}{\omega^2-c_0^2(1-4 x)q^2+i0}.
\end{eqnarray}
Note that only the coefficient in front of $q^2$ is changing with doping, the
$\omega$-term is not changed since the doping correction comes from 
very high energy fluctuations, $8t \gg \omega$.
From Eq.(\ref{gfrl1}) we deduce parameters of the effective non-linear
$\sigma$-model in the Lagrangian (\ref{eq:LL})
\begin{eqnarray}
\label{par}
&&\chi_{\perp}=\chi_{\perp}^{(0)}=1/8J, \nonumber\\
&&\rho_s=\rho_s^{(0)}(1-4 x)=\frac{J}{4}(1-4 x) \ .
\end{eqnarray}
Hence the magnon speed is reduced with the doping
\begin{eqnarray}
\label{swv}
c=\sqrt{\frac{\rho_s}{\chi_{\perp}}}=c_0\sqrt{1-4 x}.
\end{eqnarray}
Softening of magnons in the energy interval $70meV < \omega < 200meV$ which we call
the ``inermediate regime'' was observed in inelastic neutron scattering \cite{Bourges1997,Lipscombe09,Vignolle07}.
To illustrate this in the left panel of Fig.\ref{soft} we present data for LSCO
from  Refs.\cite{Coldea2001,Lipscombe09,Vignolle07} for doping levels $x=0$, $x=0.085$, and $x=0.16$.
In the right panel of Fig.\ref{soft} we plot the theoretical dispersion (\ref{swv})
for the same values of doping. The agreement between theory and experiment
in the ``intermediate regime'' is remarkable even at $x=0.16$
where the spin wave velocity reduction is approximately $40\%$.

Physics that we have discussed in the present section concerns relatively high energies 
and it is irrelevant to Lifshitz point magnetic criticality.
So, the doping dependence presented in Eqs.(\ref{par}) and (\ref{swv}) is only weakly
sensitive to $\lambda$. In the interval (\ref{lll}) it is practically $\lambda$-independent.

We have calculated the doing dependence of $c(x)$ and $\rho_s(x)$.
What can we say about doping dependence of other parameters of the Lagrangian
(\ref{eq:LL})? Inverse effective masses $\beta_1$ and $\beta_2$ are slightly dependent 
on doping. The doping icreases $\beta_1$ and decreases $\beta_2$ in such a way that the
average effective inverse mass $\beta=\sqrt{\beta_1\beta_2}$ is approximately doping 
independent \cite{Sushkov04a}.
Here we disregard the weak doping dependence of $\beta_1$ and $\beta_2$.
On the other hand, doping dependence of the coupling constant $g$ is expected to be significant.
Due to the magnon softening (\ref{swv}) the coupling constant $g$
must be decreasing with doping. Unfortunately we do not know how to perform
a reliable calculation of the coupling constant reduction with doping.
In what follows we will expect that $g$ varies with doping in such a way that
the magnetic criticality parameter $\lambda$ defined by Eq.(\ref{ll}) is approximately doping
independent.

\section{ Magnetic criticality at the Lifshitz point and the spin liquid phase}\label{sec:phase_diagr}

We start our analysis of the low energy ``infrared regime''+''intermediate regime'' from the usual collinear AFM state.
The staggered magnetization is directed along the $z$ axis and we use the standard representation (\ref{gfrl}).
The dynamics is described by the effective Lagrangian (\ref{eq:LL}).
To explain our idea we first consider circular Fermi pockets and then consider ellipticity of the pockets.
This is a conceptual section, so we derive general equations, but perform specific calculations
only for very small doping $x$ where the calculations can be done analytically with logarithmic accuracy.
\subsection{Circular Fermi pockets, $\beta=\beta_1=\beta_2$}\label{sec:circ_pock}
 The magnon Green's function reads
\begin{equation}\label{eq:D}
D(|\omega|<\Lambda,q<\Lambda_q) = \frac{\chi_{\perp}^{-1}}{\omega^2 - c^2q^2 - P^F(\omega,\bm q) + i0},
\end{equation}
Formally this equation is similar to Eq.(\ref{gfrl}), and the polarization operator is given by
the standard loop diagram, as shown in Fig. \ref{hGF1}b.
However, there are two important differences. (i) Unlike Eq.(\ref{gfrl}) which contains the bare magnon speed
$c_0$,  Eq.(\ref{eq:D}) contains the renormalized  magnon speed $c$ given by Eq.(\ref{swv}). 
(ii) Eq.(\ref{gfrl}) contains the ``ultraviolet'' polarization operator $P^{\Lambda}$ which results from
the ``coherent-incoherent'' contribution and from the energy scale up to $24J \sim 3$eV.
On the other hand Eq.(\ref{eq:D}) contains the ``coherent-coherent'' polarization operator $P^{F}$ which
comes from the low energy fluctuations at the scale $\epsilon \sim \epsilon_F \sim 20-50$meV.

The normalization point of our field theory is $q=\Lambda_q$, $\omega=c_0\Lambda_q$.
At the normalization point ${\vec n}=(0,0,1)$, i.e ${\vec n}_{\perp}=0$.
The quantum transversal fluctuation (we assume zero temperture) of the staggered magnetization
at $q=\omega=0$ reads
\begin{eqnarray}
\label{np}
\langle \vec n_{\perp}^2\rangle = -2\sum_{q<\Lambda_q}\int_{|\omega|<c_0\Lambda_q} \frac{d\omega}{2\pi i}\: D(\omega,\bm q) \ .
\end{eqnarray}
The factor $2$ comes from summation over magnon transverse polarizations.
The magnon polarization operator $P^F$ reads (see Ref. \cite{Milstein08}),
\begin{eqnarray}
\label{eq:PF}
&&P^F(\omega,\bm q) = \frac{2}{\chi_\perp} \sum_{k,\alpha} f^{\alpha}_k(1-f^\alpha_{k+q})\times \nonumber\\ 
&&\frac{ \left[ \sqrt{2}g(\bm e_\alpha \cdot \bm q) \right]^2 }{ \omega + \epsilon_k^\alpha - \epsilon_{k+q}^\alpha + i0 } 
+  \{\omega\rightarrow -\omega,\, \bm q \rightarrow -\bm q\} \ .
\end{eqnarray}
Here $f_k^\alpha = \theta(\epsilon_F - \epsilon^\alpha_k)$ denotes the zero temperature Fermi-Dirac distribution 
for holons in the pocket $\alpha$.
The expression in the brackets $[\ldots]$ is the fermion-magnon vertex that follows from the bottom line in Lagrangian (\ref{eq:LL}).
Eq. (\ref{eq:PF}) up to the prefactor is the usual 2D Lindhard Function.
The prefactor $\propto \bm q^2$ is dictated by the Adler's theorem.
After the  Wick rotation from real to imaginary frequency, $\omega=i\xi$, the polarization operator reads
\begin{equation}
\label{eq:P_magn}
P^F(i\xi,\bm q) = - \lambda c^2 q^2 \left( 1 - \frac{2}{q^2} \textrm{Re} \sqrt{ \left( \frac{ q^2}{2} + i \frac{\xi}{\beta} \right)^2 - p_F^2 q^2 } \right).
\end{equation}
Here we assume quadratic holon dispersion, $\epsilon_k^\alpha = \beta \bm p^2/2$ ($\bm p = k - \bm k_0^\alpha$),
the  Fermi momentum is $p_F = \sqrt{\pi x}$.
Since natural scales in (\ref{eq:P_magn}) are $\epsilon_F$ and $p_F$,
it is convenient to express the polarization operator in terms of dimensionless energy and momentum
\begin{eqnarray}
\label{til}
{\tilde q}=\frac{q}{p_F},\quad{\tilde \xi}=\frac{\xi}{\epsilon_F} \ .
\end{eqnarray}
Hence the quantum fluctuation (\ref{np}) reads
\begin{eqnarray}
\label{np1}
&&\langle \vec n_{\perp}^2\rangle =\frac{\beta x}{2\pi\rho_s}\int_0^{\Lambda/p_F}d{\tilde q}F({\tilde q}),\\
&&F({\tilde q})={\tilde q}\int_0^{c\Lambda/\epsilon_F}
\frac{d{\tilde \xi}}{\gamma {\tilde \xi}^2+{\tilde q}^2(1-\lambda r)},\label{eq:Fq}
\end{eqnarray}
where $\gamma=\frac{\pi\beta^2}{4c^2}x \ll 1$ and
\begin{eqnarray}
\label{po}
&&r=Re\left\{1-\frac{1}{{\tilde q}^2}\sqrt{({\tilde q}^2+i{\tilde \xi})^2-4{\tilde q}^2} \right\}.
\end{eqnarray}
We consider the collinear phase, hence $\lambda < 1$. 
The central point is that the integral (\ref{np1}) is logarithmically diverging in the limit $\lambda\rightarrow 1$.
The main contribution to the integral comes from very small $\xi$ where
 the function $r$ in Eq.(\ref{po}) can be expanded as
\begin{eqnarray}
\label{po1}
r\approx 1-\frac{|\tilde\xi|}{{\tilde q}\sqrt{4-{\tilde q}^2}}.
\end{eqnarray}
Evaluation of the $\tilde\xi$ integral in Eq.(\ref{eq:Fq})
results in 
\begin{eqnarray}
\label{FF}
F({\tilde q})=\theta(4-{\tilde q}^2)\sqrt{4-{\tilde q}^2}\ln\left(\frac{1}{1-\lambda}\right)+
f({\tilde q}) \ ,
\end{eqnarray}
where $f({\tilde q})$ only weakly dependends on $\lambda$.
The ${\tilde q}$-integration in (\ref{np1}) is straightforward
\begin{eqnarray}
\label{np2}
&&\langle \vec n_{\perp}^2\rangle=\frac{\beta x}{2\rho_s} \ln\left(\frac{1}{1-\lambda}\right)
+\phi(\lambda, \gamma) \ ,
\end{eqnarray}
where again $\phi(\lambda, \gamma)$ is non-singular and only weakly depends on  $\lambda$ near $\lambda \approx 1$. 
In the limit $x\rightarrow 0$ we should recover the result for the 2D Heisenberg model, hence $\langle \vec n_\perp^2 \rangle = \phi(\gamma\rightarrow 0, \lambda) \approx  0.8$. In the rest of the Section we will assume that $\phi(\gamma,\lambda)\approx \phi$ is approximately constant.
The logarithmic singularity in Eq. (\ref{np2}) at $\lambda\rightarrow \lambda_{LP} = 1$ is of central importance.
The singularity indicates an instability of the AFM state when $\lambda$ is sufficiently close to unity.
The singularity is similar to the logarithmic divergence of transversal spin fluctuations in Ioffe-Larkin at the Lifshitz point in frustrated magnets \cite{Ioffe88,Kharkov18}.
In addition, the singluarity is also analogous to the logarithmic divergence in 2D Heisenberg model at finite temperature \cite{takahashi1987}.
The doping $x$ in this case plays a role of an effective temperature.
The divergence indicates the quantum phase transition to the disordered spin liquid phase.

There is a critical value of the fluctuation 
\begin{eqnarray}
\label{ncc}
\langle \vec n^2_\perp \rangle_c\sim 1
\end{eqnarray}
 that is sufficient to destroy the long range AFM order. This is a sort of Lindemann
criterion for quantum melting. We will discuss value of  $ \langle \vec n^2_\perp \rangle_c$
later. Now we consider the problem conceptually.
To find the critical value $\lambda_{c1}<1$ for transition to the spin liquid phase we 
only need to equate the right hand side of Eq.(\ref{np2}) to $\langle \vec n^2_\perp \rangle_c$.
This  gives
\begin{eqnarray}
\label{lc1a}
1 - \lambda_{c1} \propto \exp{\left[-\frac{2\rho_{s}( \langle \vec n^2_\perp \rangle_c - \phi)}{\beta x}\right]}.
\end{eqnarray}
Formula (\ref{lc1a}) determines the left boundary on the phase diagram Fig.\ref{F1}b.
Note, that Eq.(\ref{lc1a}) is valid only at very small $x$. For realistic $x$ one 
needs a numerical calculation performed later.

In the spin liquid phase at $\lambda > \lambda_{c1}$ the magnon gap $\Delta$ is 
opened \cite{Ioffe88,Kharkov18}
and hence the Green's function (\ref{eq:D}) is transformed to
\begin{equation}
\label{eq:Dx}
D(i\xi,\bm {q}) = -\frac{\chi_{\perp}^{-1}}{\xi^2 + c^2q^2 + \Delta^2 + P^F(i\xi,\bm q)}.
\end{equation}
In essence $\Delta$ is the Lagrange multiplier $\mathcal{L} \rightarrow \mathcal{L} + \Delta^2(\vec n^2_{\perp} - \langle \vec n^2_{\perp} \rangle_c)$ 
that has to be determined from the condition
\begin{equation}
\label{eq:gap_cond}
\langle \vec n^2_\perp \rangle_c=
2\sum_{|\bm q|<\Lambda_q} \int_{|\xi|\leq c\Lambda_q} \frac{d \xi}{2\pi} \frac{\chi_\perp^{-1}}{\xi^2 
+ c^2 \bm q^2 + \Delta^2 + P^F(i\xi, \bm q)} \ .
\end{equation}
By construction the gap $\Delta$ vanishes at $\lambda=\lambda_{c1}$.

It is instructive to calculate $\Delta$ exactly at the Lifshitz point, $\lambda=1$.
Performing calculations by analogy with to Eqs.(\ref{np1})-(\ref{np2}) one finds with logarithmic accuracy 
\begin{eqnarray}
\label{gap11}
{\Delta}_{\lambda=1}(x) \propto
\exp{\left[-\frac{\rho_s (\langle \vec n^2_\perp \rangle_c - \phi)}{\beta x}\right]}.
\end{eqnarray}
We stress again that here we assume the limit of very small $x$. For realistic $x$ we will 
perform numerical calculation in Section \ref{uv}.

Now we consider the case $\lambda > 1$. At at a fixed doping $x$ and at a
sufficiently large $\lambda=\lambda_{c2}>1$  the spin liquid phase becomes unstable towards condensation 
of static spin spiral, see phase diagram in Fig.\ref{fig:RIXS_magnon}b.
The instability manifests as a pole in the Green's function (\ref{eq:Dx}) at $\xi=0$.
At $\xi =0$ the denominator in (\ref{eq:Dx}) is 
\begin{eqnarray}
\label{den}
D^{-1}(0,q) \propto \Delta^2+c^2q^2\left[1-\lambda\left(1-Re\sqrt{q^2-4p_F^2}\right)\right].
\end{eqnarray}
The inverse propagator has a  minimum at $q=2p_F$. Hence the  instability of the
spin liquid with respect to the static spin spiral condensation is determined from the condition
that the denominator of the magnon propagator equal to zero at $q=2p_F$,
\begin{eqnarray}
\label{zf}
\Delta^2 - 4c^2p_F^2(\lambda-1)=0 \ .
\end{eqnarray}
The critical line $\lambda_{c2}$ can be found by solving Eq. (\ref{zf}) together with Eq.(\ref{eq:gap_cond}).
Solving these Eqs. in logarithmic approximation at very small $x$ we find 
\begin{eqnarray}
\label{lc2}
\lambda_{c2}-1 \propto \exp{\left(-\frac{2\rho_s (\langle \vec n^2_\perp \rangle_c - \phi)}{\beta x}\right)}.
\end{eqnarray}
There are three points to note.
(i) While $\Delta$ is zero at the left borderline
of the spin liqud phase, $\lambda=\lambda_{c1}$, the gap
is nonzero at the right borderline $\lambda=\lambda_{c2}$, see Fig.\ref{fig:RIXS_magnon}b.
However, in this case $\Delta$ is not the real magnon gap, the magnetic pseudogap corresponds to the distance from the real $\omega$-axis to the nearest pole in magnon's Green's function.
Since at the phase boundary $\lambda_{c2}$ the magnon's Green's function acquires a pole at zero frequency, the spin excitation gap is zero in agreement with the Goldstone theorem. 
(ii) At $\lambda > \lambda_{c2}$  the static spin-spiral with the wave vector $Q=2p_F$ condenses
\begin{eqnarray}
\label{str}
{\vec n}=A[\vec{e_1}\cos({\bf Q}\cdot{\bf r}) + \vec {e_2} \sin({\bf Q}\cdot{\bf r})] \ .
\end{eqnarray}
Close to the phase transition line the amplitude $A$ is very small.
(iii) Direction of the spiral wave vector ${\bf Q}$ can be  arbitrary.  This is because
for circular Fermi pockets considered in this subsection
 our field theory ``does not know'' about the lattice orientation.

\subsection{Elliptic Fermi pockets}\label{sec:ell_pock}
In order to describe a situation relevant to cuprates,
we consider elliptic Fermi pockets stretched along the face of the MBZ, see Fig.\ref{FS1}.
We still use the parabolic approximation, the second line in Eq.({\ref{disp}), $\beta_1> \beta_2$
The magnon polarization operator in the case of elliptic pockets could be obtained from (\ref{eq:P_magn}) 
by performing rescaling of $q_{1,2}$ in Eq.(\ref{eq:PF}).
Hence the dimensioneless polarization operator $r(i\xi,\bm q)$ in Eqs. (\ref{eq:Fq}) and (\ref{po1}) should be 
replaced by
\begin{eqnarray}
\label{eq:r_ellipt}
&&q^2\lambda r \to \lambda(q_1^2r_a+q_2^2r_b)=\lambda p_F^2({\tilde q}_1^2r_a+{\tilde q}_2^2r_b)\nonumber\\
&&r_{\mu=\{a,b\}}=Re\left\{1-\frac{1}{{\tilde q}_\mu^2}\sqrt{({\tilde q}_\mu^2+i{\tilde \xi})^2-4{\tilde q}_\mu^2} \right\}
\end{eqnarray}
where the effective Fermi momentum  $p_F=\sqrt{\pi x}$ remains the same and we  define 
\begin{eqnarray}
{\tilde q}_a^2=\sqrt{\frac{\beta_1}{\beta_2}}{\tilde q}_1^2+\sqrt{\frac{\beta_2}{\beta_1}}{\tilde q}_2^2, \quad 
{\tilde q}_b^2=\sqrt{\frac{\beta_1}{\beta_2}}{\tilde q}_2^2+\sqrt{\frac{\beta_2}{\beta_1}}{\tilde q}_1^2.
\end{eqnarray}

The calculation of the N\'eel - spin liquid phase boundary line $\lambda_{c1}$ is analogous to the case of circular Fermi pockets
presented in the Section  \ref{sec:circ_pock}.
Eq.(\ref{lc1a}) is replaced by
\begin{eqnarray}
\label{lc1}
1 - \lambda_{c1} \propto \exp{\left[-\frac{2\rho_{s}(\langle \vec n^2_\perp \rangle_c - \phi)}{\beta x f(\beta_1, \beta_2)}\right]}\ ,
\end{eqnarray}
where $f(\beta_1, \beta_2)$ is a smooth symmetric function that only weakly depends on the 
ratio $\beta_1/\beta_2$, for circular pockets $f(\beta,\beta)=1$. So here the ellipticity does not
result in a significant effect.
 
Importantly, for $\lambda > 1$ the ellipticity results in a qualitative effect.
It pins the wave vector of the spin spiral to the antinodal direction, $\bm Q = (Q,0)$ or $\bm Q = (0,Q)$.
To see this one has again to write down the denominator of the magnon Green's function in the spin liquid phase,
similar to Eq.(\ref{den}), but with an account of anisotropic polarization operator (\ref{eq:r_ellipt}).
Then for the nodal direction, $\bm Q = Q/\sqrt{2}(1, \pm 1)$, the denominator  has a minimum at a
\begin{eqnarray}
\label{Qnod}
Q=2p_F\left(\frac{\beta_2}{\beta_1}\right)^{1/4} \ ,
\end{eqnarray}  
and the instability  condition (\ref{zf}) is replaced by
\begin{eqnarray}
\label{zfnod}
\Delta^2 - 4c^2p_F^2\sqrt{\frac{\beta_2}{\beta_1}}(\lambda-1)=0 \ .
\end{eqnarray}
On the other hand for the antinodal direction, $\bm Q = (Q,0)$ or $\bm Q = (0,Q)$, 
the denominator has a minimum at the wave vector
\begin{eqnarray}
\label{Qanod}
Q=\frac{2p_F}{\sqrt{\frac{1}{2}\left(\sqrt{\frac{\beta_1}{\beta_2}}+\sqrt{\frac{\beta_2}{\beta_1}}\right)}} \ ,
\end{eqnarray}  
and the instability  condition reads
\begin{eqnarray}
\label{zfanod}
\Delta^2 - 4c^2p_F^2(\lambda-1)\frac{2}{
\sqrt{\frac{\beta_1}{\beta_2}}+\sqrt{\frac{\beta_2}{\beta_1}}}=0  \ .
\end{eqnarray}
This condition is satisfied at a smaller value of $\lambda$ than the diagonal spin spiral condition (\ref{zfnod}).
Hence the spin spiral always condensates in the antinodal direction.
The wave vector is given by Eq.(\ref{Qanod}), but this Eq. is valid only at very small $x$.

In spite of the pinning of the spin spiral direction the spin liquid borderline is not changed much
compared to Eq.(\ref{lc2}). Taking into account the Fermi pocket ellipticity the equation for the critical line reads
\begin{equation}
\lambda_{c2} - 1 \propto \exp\left( - \frac{2\rho_s (\langle \vec n^2_\perp \rangle_c - {\phi}) }{\beta x f(\beta_1, \beta_2)} \right).
\end{equation}

\section{Ultraviolet cutoff, ``Lindemann criterion'',  numerical calculations and comparison with experiment} \label{uv}
\subsection{Ultraviolet cutoff}
Let us first determine $\Lambda_q$. At zero doping, $x=0$, Eq.(\ref{np}) reads
\begin{eqnarray}
\label{np0}
\langle \vec n_{\perp}^2\rangle = \frac{2}{\chi_{\perp}}\int_0^{\Lambda_q}\frac{d^2q}{(2\pi)^2}
\int_{-c_0\Lambda_q}^{c_0\Lambda_q} \frac{d\xi}{2\pi }\  \frac{1}{\xi^2+c_0^2q^2} \ .
\end{eqnarray}
This corresponds to the usual 2D Heisenberg model on the square lattice where we know well  that the staggered magnetization is
$\langle S_z\rangle\approx 0.30$.
This corresponds to $\langle 2S_z\rangle=\langle n_z\rangle\approx 1-\frac{1}{2}\langle \vec n_{\perp}^2 \rangle\approx0.6$.
Hence at $\langle \vec n_{\perp}^2 \rangle=0.8$. The upper limit of integartion in Eq.(\ref{np0}) must be tuned
to reproduce this value. From here we find
\begin{eqnarray}
\label{lamq}
\Lambda_q \approx 1.2=0.19(r.l.u) \ .
\end{eqnarray}
This is the value of $\Lambda_q$ for crossover from the ``intemediate regime'' to the ``ultraviolet regime'' that was
first introduced in Fig.\ref{fig:RIXS_magnon} based on experimental data.

\subsection{``Lindemann criterion''}
The concept of the critical value of magnetic fluctuation is defined by Eq.(\ref{ncc}).
Here we quantify the value of $\langle \bm n^2_\perp \rangle_c$, the ``Lindemann criterion''.
This value depends on dimensionality and probably on some details of fluctuations. 
In Ref.\cite{Kharkov18} comparing the field theory with numerical DMRG data
we found that for 1D integer spin Haldane chain $\langle \bm n^2_\perp \rangle_c\approx 0.6$.
Interestingly, the renormalization group in this case gives $\langle \bm n^2_\perp \rangle_c=1$,
Ref.\cite{Affleck89}, although DMRG is more reliable.
In the same paper\cite{Kharkov18} we argue that for 2D Ioffe-Larkin spin liquid
\begin{eqnarray}
\label{nccL}
\langle \bm n^2_\perp \rangle_c\approx 1 \ .
\end{eqnarray}

Here we would like also make a comparison of our approach with Takahashi's modified spin wave theory\cite{takahashi1987}
or Schwinger boson mean field technique\cite{Auerbach88}.
For  2D Heisenberg model at nonzero temperature $T$ these methods work reasonably well.
 In this case  equation similar to (\ref{np}) reads
\begin{eqnarray}
\label{npT}
&&\langle \vec n_{\perp}^2\rangle_c = \frac{2}{\chi_{\perp}}\int_0^{\Lambda_q}\frac{1}{\omega_q}
\left(\frac{1}{e^{\omega_q/T}-1}+\frac{1}{2}\right)\frac{d^2q}{(2\pi)^2} \ .
\end{eqnarray}
Here $\omega_q=\sqrt{c^2_0q^2+\Delta_T^2}$, $\Delta_T$ is the temperature related ``gap''.
At $T\ll J$  we can rewrite (\ref{npT}) as
\begin{eqnarray}
\label{npT1}
\langle \vec n_{\perp}^2\rangle_c = \frac{4T}{\pi J}\ln\left(\frac{T}{\Delta_T}\right)
+\frac{2\sqrt{2}}{\pi}\int_0^{\Lambda_q}dq.
\end{eqnarray}
The second term in this equation is the zero temperature quantum fluctuation and according to
the discussion in the previous paragraph this term is approximately equal to $0.8$.
On the other hand according to Ref.\cite{takahashi1987} the finite $T$  gap is  
$\Delta_T \sim T e^{-2\pi\rho_s^{(r)}/T} $, where $\rho_s^{(r)}\approx 0.17 J$ is the renormalized spin stiffness for the 2D Heisenberg model.
Substitution of $\Delta_T$ in Eq. (\ref{npT1}) gives $\langle \vec n_{\perp}^2\rangle_c \approx 2$.
This is the expected result since the Takahashi's modified spin wave theory implicitly assumes
the leading order expansion $\langle n_z\rangle =\langle \sqrt{1-\vec n^2_{\perp}}\rangle\approx
1-\frac{1}{2}\langle\vec n^2_{\perp}\rangle$ and equating $\langle n_z\rangle$ to zero.
This immediately gives the above condition.
The value of $\langle\vec n^2_{\perp}\rangle > 1$ looks strange keeping in mind the constraint $n^2=1$.
The large fluctuation is a byproduct of linearization which one necessarily does when working with
strong fluctuations. It is known that for the 2D Heisenberg model at $T\ne 0$ the method works 
reasonably well for the correlation length\cite{Manousakis1991}.
However, when applied to a disordered system at zero temperature  the method gives strange result
that the physical gap is $2\Delta_T$, Ref.\cite{Auerbach88} 
In this case $\Delta_T$ is just an infrared cutoff unrelated to temperature.
Moreover, application of the criterion $\langle \bm n^2_\perp \rangle_c=2$ to the Ioffe-Larkin spin liquid
in 2D $J_1-J_3$  model and to 1D Haldane spin chain gives results completely inconsistent with 
numerics \cite{Kharkov18}. Therefore in the present work we use the criterion (\ref{nccL}).

\subsection{Numerical calculations and comparison with experimental data}\label{sec:num_phase_d}
As soon as the ultraviolet cutoff (\ref{lamq}) and the quantum melting criterion (\ref{nccL}) are fixed we can
find the phase digarm by solving numerically Eqs.(\ref{eq:gap_cond}) and (\ref{zfanod}).
The phase diagram resulting from this calculation is presented in Fig.\ref{PDC}.
\begin{figure}
\includegraphics[scale=.3]{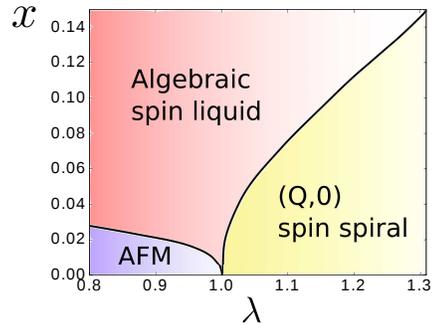} 
\caption{Zero temperature $\lambda-x$ phase diagram of the modified $t-J$ model.
The range of $\lambda$ corresponding to cuprates is given by Eq.(\ref{lll}).
}
\label{PDC}
\end{figure}

It is even more instructive to calculate the ``gap'' $\Delta$ defined by Eq.(\ref{eq:Dx}).
The ``gap'' is determined from a numerical solution of Eq.(\ref{eq:gap_cond}).
At $\omega=\Delta$ the magnetic response is maximum at $q=0$ which for neutron scattering
corresponds to ${\bf q}=(\pi,\pi)$. Therefore $\Delta$ is identical to $E_{cross}$ 
usually determined in neutron scattering\cite{Fujita12}.
The value of $\Delta$ depends on $\lambda$ at a given doping $x$, for instance it vanishes at the transition line from the spin liquid to the AFM phase.
However, at $\lambda > 1$ the dependence $\Delta(\lambda)$ is rather weak, this even includes the
transition line from the spin liquid to the spin spiral state. In this region it is sufficient to calculate  
$\Delta(\lambda=1) \approx E_{cross}$. The result of this calculation is shown in Fig.\ref{Ecross}
by the black solid line.
\begin{figure}
\includegraphics[scale=.32]{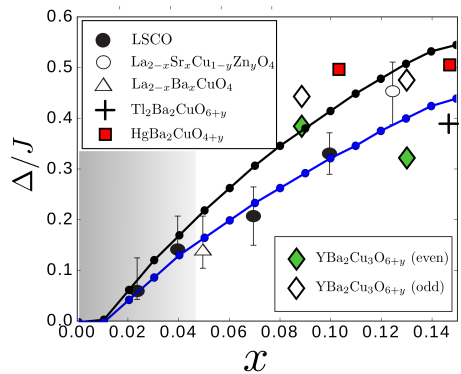}
\caption{$E_{cross}$ versus doping. 
The black solid line shows the calculation without account of the
momentum dependence of the holon residue. 
The blue solid line shows the calculation with account 
 the holon residue momentum dependence.
The range between these lines indicates the theoretical
uncertainty of our calculation.
Experimental data are shown by symbols \cite{Fujita12, He02, Pailhes06, Chan16, Chan16a}.
Theoretical curves correspond to $\lambda=1$, however, dependence of $E_{cross}$ on $\lambda$ is weak. 
Symbols at $x<5\%$ corresponds to LSCO (grey region), where physics is driven by localization. Comparison of theoretical calculations and experimental data at doping $x<5\%$ is less justified.
}
\label{Ecross}
\end{figure}
There is an uncertainty in our calculations that is worth mentioning.
In our calculation we assume that the coupling constant $g$ defined 
in Eq.(\ref{eq:LL}) is momentum independent.
Within the $t-J$ model the coupling constant is $g=t\sqrt{Z_{\bm k}Z_{\bf k+q}}$.
At small $q$ we obtain $g=t Z$, where $Z=Z_{(\pi/2,\pi/2)}$.
At doping $x \sim 0.1$ the Fermi momentum $p_F =\sqrt{\pi x}\sim 0.6$
and the typical value of momentum responsible for fluctuations, $q\sim 2p_F$,
is quite large. At these values of momentum the dependence of the quasiparticle residue on momentum
becomes significant. Fitting numerical data obtained in SCBA we found that
\begin{equation}
\label{eq:Z_q}
 Z_{\bm k} \approx Zz_{\bm k}\ , \ \ \  z_{\bm k}= 1 - 0.3\frac{\epsilon_{\bm k}}{J}\ ,
 \end{equation}
where $\epsilon_{\bm k}$ is given by Eq.(\ref{disp}).
Within the range of parameters corresponding to cuprates the coefficient in the fit varies
between 0.25 and 0.35. We take 0.3 as some effective value.
The fit (\ref{eq:Z_q}) is valid when $Z_k\geq 0$, otherwise $Z_k=0$. 
To account for the residue momentum dependence the expression under the sum $\sum_{k,\alpha}$
in the polarization operator (\ref{eq:PF}) should be multiplied by $z_{\bm k}z_{\bm k+q}$ and all other formulas are unchanged.
The gap $\Delta$ calculated with account of $z_{\bm k}$ is plotted in Fig.\ref{Ecross} by the
blue solid line. The range between the black and the blue line indicates the theoretical
uncertainty of our calculation.
Symbols in Fig.\ref{Ecross} display experimental data.
The agreement between the theory and the experiment is exciting. Our approach grasps the essential physics of the problem.


\section{Spin-spin correlation function in the algebraic spin liquid phase} \label{sec:correl}

Here we consider the equal time spin-spin correlator,  
\begin{equation}
C(r) = \langle  {\vec n}(\bm r)\cdot {\vec n} (0) \rangle.
\end{equation}
There are two main messages of this section. (i) The correlator decays at the typical scale 
$r \sim 1/p_F \sim 1/\sqrt{x}$. This is true even in  the limit when the gap is exponentially small,
Eq.(\ref{gap11}).  (ii) There is a long distance tail of the correlator which decays as $1/r^3$, so  the 
spin liquid is algebraic.

Following Ref.\cite{Kharkov18} we represent the correlator as
\begin{eqnarray}
\label{tc}
&&C(\bm r) \approx 1+{\cal P}-R+. . . 
\end{eqnarray}
where
\begin{eqnarray}
\label{pc}
{\cal P}(r)= \langle {\bm n}_{\perp}(r)\cdot {\bm  n}_{\perp}(0)\rangle, 
\quad R = {\cal P}(0)=\langle \bm n_\perp^2 \rangle.
\end{eqnarray}
The two-point correlator is normalized such that $C(0)= 1$. In the 
spin liquid phase the correlation function should vanish at large distances, 
$C(r\rightarrow\infty)\rightarrow 0$ and ${\cal P}(r\rightarrow\infty)\rightarrow 0$. 
This condition is consistent with Eq.(\ref{tc}) if we truncate the asymptotic expansion
in Eq.(\ref{tc}) keeping only the terms explicitly presented there.
The explicit expression for ${\cal P}$ immediately follows from Eq.(\ref{eq:Dx})
\begin{eqnarray}
\label{corrp1} 
{\cal P}(r)=\frac{1}{\pi\chi_{\perp}}
\int_0^{\Lambda_q}d q \, q \, \int_0^{c\Lambda_q} \frac{d\xi}{2\pi} \frac{J_0( q r)}{\xi^2 + \Delta^2 + P^F(i\xi, q)}.
\end{eqnarray}
Here $J_0$ is the Bessel function. Note that in this section for simplicity we consider circular Fermi pockets.
At $r=0$ formula (\ref{corrp1}) is identical to Eq.(\ref{eq:gap_cond}).

\begin{figure}
\includegraphics[scale=0.3]{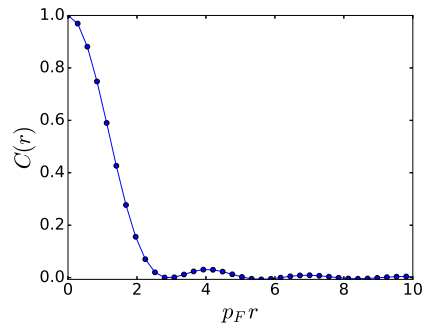}
\caption{Spin-spin equal time correlation function $C(r,t=0)=\langle \bm n (\bm r) \cdot \bm n(0) \rangle$ 
at the Lifshitz point, $\lambda=1$. Here we take circular Fermi pockets, $\beta/J=2$, doping $x=0.1$.  }
\label{fig:C(r)}
\end{figure}

The algebraic behaviour of the correlator originates from the nonanalytic dependence on $\xi$ in  the
polarization operator Eq.(\ref{po1}).
Here we calculate the correlator at the Lifshitz point, $\lambda=1$.
Evaluation of the $\xi$ integral in (\ref{corrp1}) in the limit $\Delta \to 0$ with logarithmic accuracy gives
\begin{eqnarray}
\label{corrp2} 
{\cal P}(r)&=&\frac{\beta}{2\pi^2\rho_s}
\int_0^{2p_F}dq \sqrt{4p_F^2-q^2}\nonumber\\
&\times&\ln\left(1+\frac{c^2}{\Delta^2}q \sqrt{4p_F^2-q^2}\right)J_0( q r).
\end{eqnarray}
From here we come to the conclusions formulated in the beginning of the section.
{\it (i) The correlator ${\cal P}$ and hence the correlator $C$
decays at the typical scale about
$r \sim 1/p_F \sim 1/\sqrt{x}$.  (ii) There is a long distance tail of the correlator which 
decays as $1/r^3$, so  the spin liquid is algebraic.}

We evaluate the long distance asymptotics, $r\rightarrow \infty $, in Eq. (\ref{corrp2}) using the stationary phase approximation. The leading contribution to the integral comes from the endpoints of the integration, $q=0$ and $q=2p_F$. 
Performing Tailor expansion of the logarithm in (\ref{corrp2}) in the vicinity of the endpoints we obtain
that the asymptotics contains a power tail $1/r^3$ as well as the 
oscillating power tail
$\propto \frac{\cos{2p_Fr}}{r^{5/2}}$, a sort of Fridel oscillations.
The $1/r^3$ asymptotics is due to the left endpoint $q=0$  and the oscillating part of the asymptotics is due to the right endpoint $q=2p_F$ of the integration.
However, the oscillating power tail is present only in the case of the circular Fermi pockets. In fact,
the oscillations are strongly suppressed for elliptic pockets (some algebraically decaying oscillations survive for the
nodal direction), while the points (i) and (ii) are generic.
Equation (\ref{corrp2}) is valid only at a very small doping $x$ where the logarithmic approximation
makes sense. On the other hand numerical integration in Eq.(\ref{corrp1}) is straightforwrd.
The correlator $C(r)$  calculated by performing numerical integration at $x=0.1$ in Eq. (\ref{corrp1}) is plotted in Fig.\ref{fig:C(r)}.

\section{Conclusion}\label{sec:discuss}
In the present work we demonstrate that there is a hidden dimensionless parameter $\lambda$ which drives quantum magnetic
criticality in the  extended $t-J$ model at low doping $x$. Using an effective field theory  we study the zero temperature 
$\lambda-x$ phase diagram of the model. The phase digram is shown schematically in Fig.\ref{F1}b and 
quantitatively in Fig.\ref{PDC}. The most important feature of the phase diagram is the quantum tricritical Lifshitz point at $x=0$, $\lambda=1$.
We calculate parameters of the effective theory using a self consistent Born approximation.
Using this approximation we show that underdoped cuprates are close to the quantum tricritical point.
The three phases ``meet'' at the tricritical point: N\'eel antiferromagnet,  Spin spiral
with antinodal direction of the spiral wave vector and algebraic spin liquid. 
We believe that underdoped cuprates belong either to the spin 
liquid phase or they are on the borderline between the spin liquid and the antinodal spin spiral.
We study properties of the spin liquid phase and demonstrate  algebraic decay of equal time 
 spin spin correlation. We calculate the energy position $E_{cross}$ of the inelastic neutron scattering response
maximum at ${\bm q}=(\pi,\pi)$ and compare our results with experiments. 
Theoretical curves and experimental data are displayed in Fig.\ref{Ecross}.
We also explain softening of magnons
in the intermediate regime observed in inelastic neutron scattering, see Fig. \ref{soft}.

\section{Acknowledgments}
The work has been supported by Australian Research Council No DP160103630.

\end{document}